\documentclass[]{aa}  

\usepackage{graphicx}
\usepackage{txfonts}

\newcommand{\ltsima} {$\; \buildrel < \over \sim \;$}
\newcommand{\gtsima} {$\; \buildrel > \over \sim \;$}
\newcommand{\lta} {\lower.5ex\hbox{\ltsima}}
\newcommand{\gta} {\lower.5ex\hbox{\gtsima}}

\newcommand{\kms}{km\ s$^{-1}$}

\begin{document}

   \title{ALMA unveils a triple merger and gas exchange in a hyper-luminous radio galaxy at $z$\,=\,2: the Dragonfly Galaxy (\textrm{II})}

\titlerunning{ALMA's view on the Dragonfly Galaxy}

   \author{B.~H.~C. Emonts
          \inst{1}\thanks{Marie Curie fellow (e-mail: bjornemonts@gmail.com)}
          \and
          C. De Breuck\inst{2}
          \and
          M.~D. Lehnert\inst{3,4}
          \and 
          J. Vernet\inst{2}
          \and
          B. Gullberg\inst{2}
          \and
          M. Villar-Mart\'{i}n\inst{1,5}
          \and
          N. Nesvadba\inst{6}
          \and
          G. Drouart\inst{7}
          \and 
          R. Ivison\inst{2,8}
          \and
          N. Seymour\inst{9}
          \and
          D. Wylezalek\inst{10}
          \and
          P. Barthel\inst{11}
          }

   \institute{Centro de Astrobiolog\'{i}a (CSIC-INTA), Ctra de Torrej\'{o}n a Ajalvir, km 4, 28850 Torrej\'{o}n de Ardoz, Madrid, Spain
        \and
             European Southern Observatory, Karl Schwarzschild Strasse 2, 85748 Garching, Germany
        \and
             CNRS, UMR 7095, Institut d'Astrophysique de Paris, 98 bis boulevard
Arago, 75014 Paris, France
        \and
             Sorbonne Universit{\'e}s, UPMC Universit{\'e} Paris VI, Institut
d'Astrophysique de Paris, 75014 Paris, France
        \and
             Astro-UAM, UAM, Unidad Asociada CSIC, Facultad de Ciencias, Campus de Cantoblanco, E-28049, Madrid, Spain
        \and
             Institut d'Astrophysique Spatiale, CNRS (UMR8617), Universit\'e Paris-Sud 11, Batiment 121, Orsay, France
        \and
            Department of Earth and Space Science, Chalmers University of Technology, Onsala Space Observatory, 43992 Onsala, Sweden
        \and
             Institute for Astronomy, University of Edinburgh, Royal Observatory, Blackford Hill, Edinburgh, EH9 3HJ, UK
        \and 
            International Centre for Radio Astronomy Research, Curtin University, Perth, Australia 
        \and
            Johns Hopkins University, Department of Physics $\&$ Astronomy, 3400 N. Charles Street, Baltimore, MD, 21218, USA
        \and
            Kapteyn Astronomical Institute, University of Groningen, 9747 AD Groningen, The Netherlands
   }

   \date{Accepted Oct.\,7, 2015}

  \abstract
   {The Dragonfly Galaxy (MRC\,0152-209), at redshift $z$\,$\sim$\,2, is one of the most vigorously star-forming radio galaxies in the Universe. What triggered its activity? We present ALMA Cycle 2 observations of cold molecular CO(6-5) gas and dust, which reveal that this is likely a gas-rich triple merger. It consists of a close double nucleus (separation $\sim$4 kpc) and a weak CO-emitter at $\sim$10\,kpc distance, all of which have counterparts in {\it HST}/NICMOS imagery. The hyper-luminous starburst and powerful radio-AGN were triggered at this precoalescent stage of the merger. The CO(6-5) traces dense molecular gas in the central region, and complements existing CO(1-0) data, which reveal more widespread tidal debris of cold gas. We also find $\sim$10$^{10}$\,M$_{\odot}$ of molecular gas with enhanced excitation at the highest velocities. At least 20$-$50$\%$ of this high-excitation, high-velocity gas shows kinematics that suggests it is being displaced and redistributed within the merger, although with line-of-sight velocities of |v|\,$<$\,500\,\kms, this gas will probably not escape the system. The processes that drive the redistribution of cold gas are likely related to either the gravitational interaction between two kpc-scale discs, or starburst/AGN-driven outflows. We estimate that the rate at which the molecular gas is redistributed is at least \.{M}\,$\sim$\,1200\,$\pm$\,500 M$_{\odot}$\,yr$^{-1}$, and could perhaps even approach the star formation rate of $\sim$3000\,$\pm$\,800\,M$_{\odot}$\,yr$^{-1}$. The fact that the gas depletion and gas redistribution timescales are similar implies that dynamical processes can be important in the evolution of massive high-$z$ galaxies.}


   \maketitle
%

\section{Introduction}

In the low-$z$ Universe, there is increasing evidence that powerful radio galaxies and ultra-luminous infrared (IR) galaxies ($L_{\rm IR} \ge 10^{12} L_{\odot}$) are often associated with gas-rich galaxy mergers \citep[e.g.,][]{hec86,ram12,san96}. In turn, massive starbursts and powerful radio jets have been observed to exert negative feedback by driving fast gas outflows \citep[e.g.,][]{hec00,mor05,mor13sci,hol08,vei13,mah13,das14,cic14,caz14,tad14,arr14,gar15}. High-$z$ radio galaxies (HzRGs; $L_{\rm 500\,MHz}>10^{27}$ W\,Hz$^{-1}$) are massive systems that are often in the regime of ultra- or hyper-luminous IR galaxies \citep[HYLIRGs, $L_{\rm IR}$\,$\ge$\,10$^{13}$\,$L_{\odot}$;][]{dro14,pod15}. In many cases these are systems where starburst and AGN activity are at their peak. To understand how processes like merging, fueling, and feedback during this epoch of peak activity drive the evolution of massive galaxies, it is vital to study the role of cold molecular gas -- the raw fuel for star formation.

Only about a dozen HzRGs have been detected in cold molecular CO-emitting gas \citep[see review by \citealt{mil08}, and more recently][]{ivi08,ivi12,nes09,emo11mnras,emo11,emo13,emo14}. From the overall CO distribution, \citet{ivi12} argue that gas-rich galaxy mergers are ubiquitous among starbursting HzRGs, while \citet{emo14} find indications of jet-induced feedback from alignments between CO emission and the radio axis. However, existing observations on cold gas and dust suffer from poor spatial resolution, making it difficult to study the detailed processes that drive the evolution of these systems.

\begin{figure}
   \centering
   \includegraphics[width=0.43\textwidth]{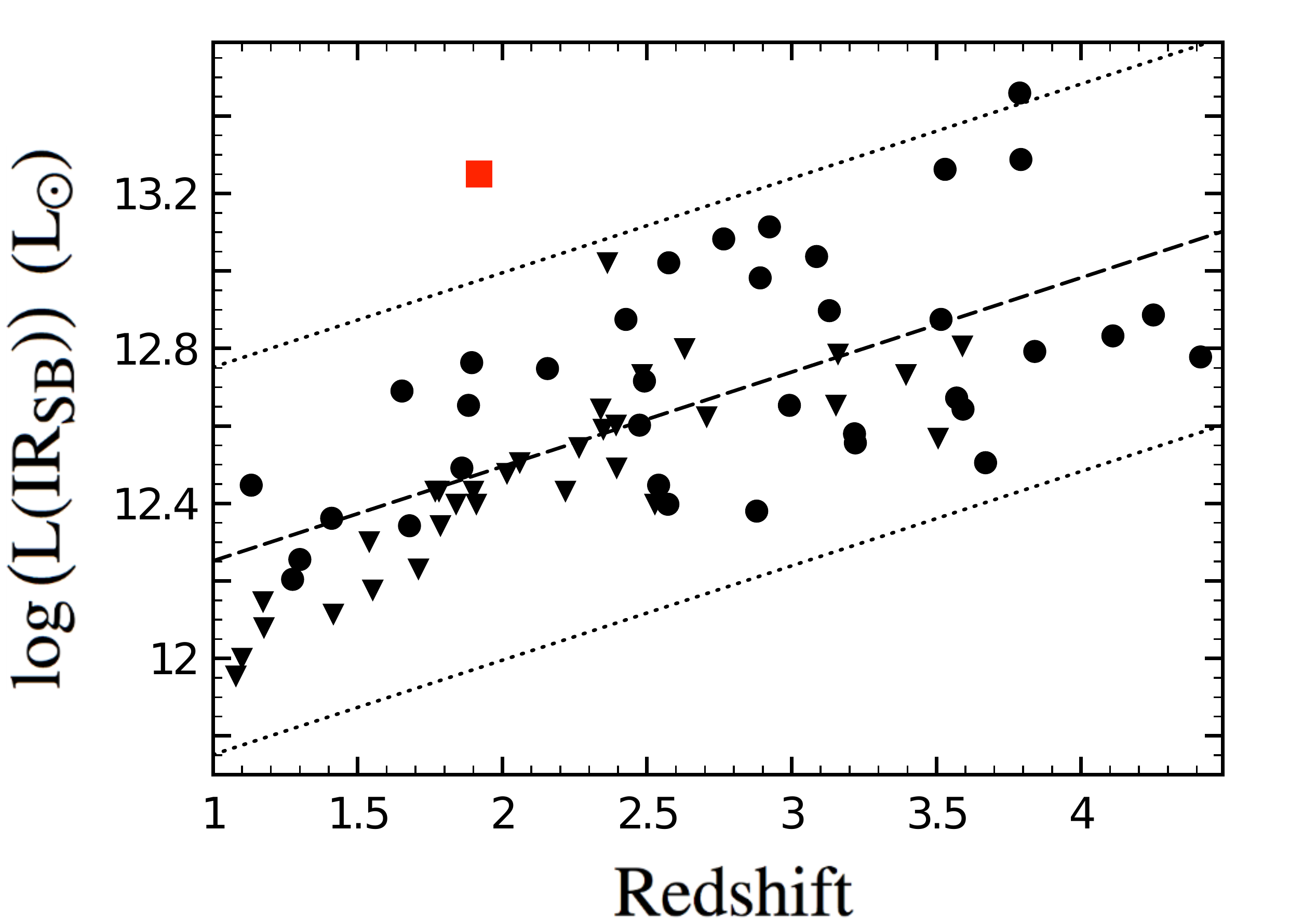}
   \caption{Starburst IR luminosity plotted against redshift for HzRGs (plot adapted from \citealt{dro14}). Any contribution from the AGN to the IR luminosity was carefully subtracted through fitting the spectral energy distribution with {\it Spitzer} and {\it Herschel} photometry \citep{dro14}. Solid circles are the IR detections, triangles the IR upper limits, and the red square is the Dragonfly Galaxy. The dashed line shows the best fit to the data, with the dotted lines $\pm$0.5 dex. The Dragonfly Galaxy is the only HzRG that deviates from the general trend by $\sim$0.8 dex. }
              \label{fig:IR}
    \end{figure}

We here present high (0.3$^{\prime\prime}$) resolution ALMA Cycle 2 observations of CO(6-5) emission and dust continuum in the Dragonfly Galaxy (MRC\,0152-209) at $z$\,=\,1.92. This is one of the most IR luminous HzRGs ($L_{\rm IR\,(SB)} \sim 2 \times 10^{13} L_{\odot}$), well in the regime of HYLIRGs, with a star formation rate determined with {\it Spitzer} and {\it Herschel} (3\,--\,500\,$\mu$m) of $\sim$\,3000 M$_{\odot}$ yr$^{-1}$ \citep{dro14}. Just how unique this source is compared to other HzRGs is visualised in Fig.\,\ref{fig:IR}, where we show that the Dragonfly Galaxy lies almost one dex above the typical starburst IR luminosity of HzRGs at similar redshifts. After initially presenting a detection of $^{12}$CO(1-0) in this system \citep[][hereafter EM11]{emo11}, we recently used the Australia Telescope Compact Array (ATCA) to image this CO(1-0), part of which appears to be widespread tidal debris from a merger \citep[][hereafter EM15]{emo15}. Our current ALMA data provide a unique view on this merger, and reveal that significant amounts of molecular gas are being redistributed within the merging system. 

Throughout this paper we assume H$_{0}$=71\,\kms\,Mpc$^{-1}$, $\Omega_{\rm M}$=0.3, and $\Omega_{\Lambda}$=0.7 (8.3 kpc\,arcsec$^{-1}$).

\section{ALMA observations}

The ALMA Cycle 2 observations in Band 6 were performed in July and Aug 2014 for a total of 6.3 min on source with 32 antennas (baselines $17 - 783$m). 
We used two 4\,GHz bands, covering 235.8 $-$239.6 and 250.1$-$254.9 GHz. The data were calibrated in CASA \citep[Common Astronomy Software Applications;][]{mcm07} with the calibration script that was supplied by the ALMA observatory. We combined both bands to image the 1.2\,mm dust continuum, excluding channels with CO(6-5) emission ($\nu_{\rm obs}$\,=\,236.7 GHz). For the CO(6-5) data, we subtracted the continuum in the UV-domain by fitting a straight line to the line-free channels. We imaged the data using various weightings \citep{bri95} and cleaned the strongest signal. For the analysis presented in this paper we used data with robustness weighting parameter of +0.5 ($0.32^{\prime\prime}$\,$\times$\,0.28$^{\prime\prime}$ beam; pa\,=\,78.9$^{\circ}$), unless otherwise indicated. The line data were binned to 30 \kms\ channels and Hanning smoothed to a velocity resolution of 60 \kms. They are presented in optical velocity definition with respect to $z$\,=\,1.9212 (EM11). The root mean square (rms) noise level is 0.13 mJy beam$^{-1}$ for the continuum and 0.54 mJy beam$^{-1}$ for the Hanning smoothed line data. All features discussed in this paper were present in both the July and Aug data. A total intensity CO(6-5) map was created by filtering the data through a low-resolution mask.

   \begin{figure}
   \centering
   \includegraphics[width=0.49\textwidth]{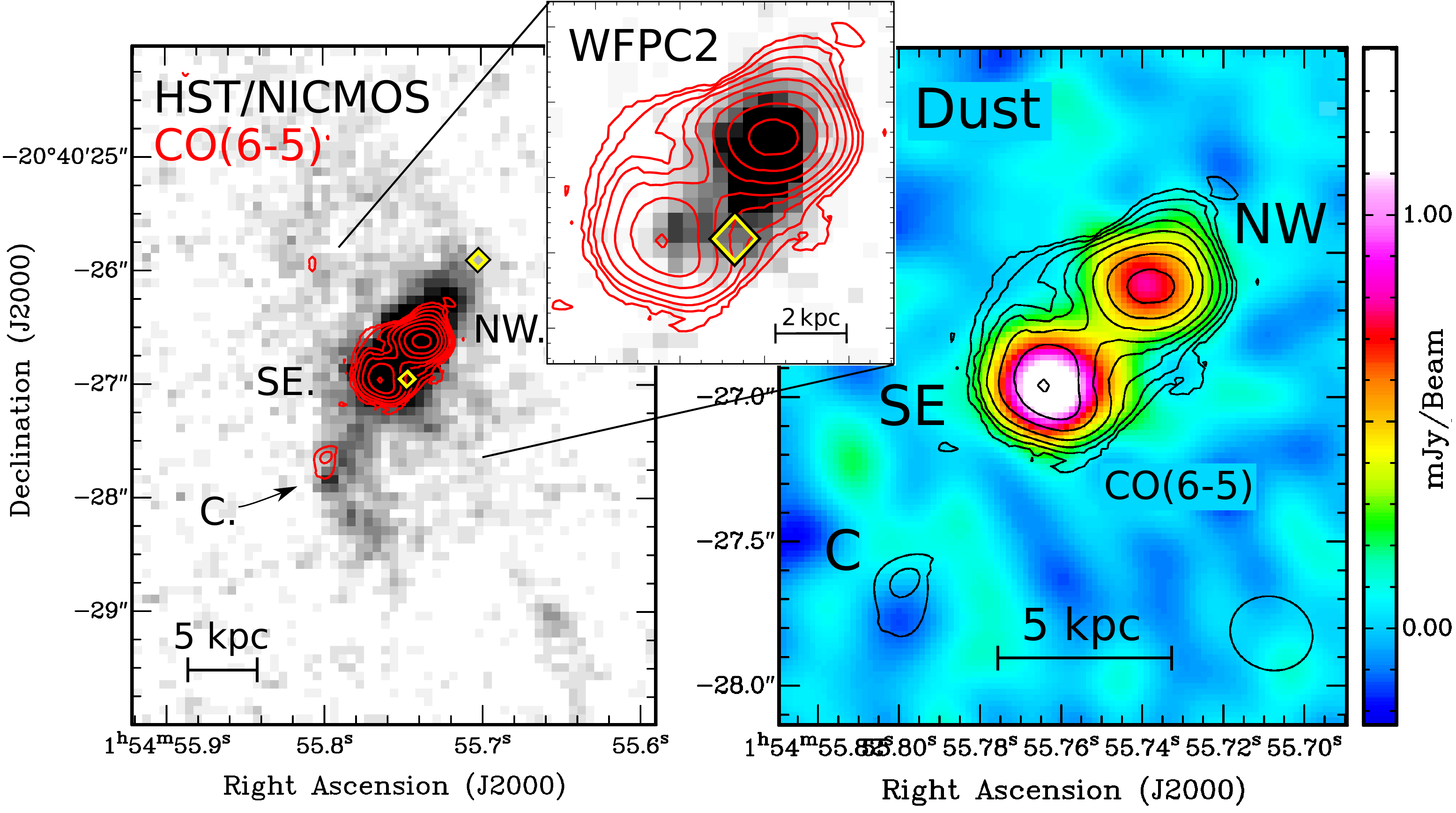}
   \caption{The Dragonfly Galaxy. {\sl Left:} Contours of CO(6-5) emission overlaid onto the {\it HST}/NICMOS F160W image from \citet{pen01}. The inset shows CO(6-5) contours of the two central components overlaid onto the {\it HST}/WFPC2 F555W image from EM15. Contour levels: 0.16\,to\,1.69\,Jy\,beam$^{-1}$\,$\times$\,\kms, in steps of factor 1.4. The {\it HST} astrometry was adjusted so that the double nucleus in the {\it HST} images matches the ALMA components to within 0.2$^{\prime\prime}$. The yellow diamonds represent the two components of the 8.2 GHz radio source \citep{pen00radio}. {\sl Right:} 1.2mm dust continuum, with the CO(6-5) contours overlaid.}
              \label{fig:morphology}
    \end{figure}

\section{Results}
\label{sec:results}

\begin{figure*}
   \centering
   \includegraphics[width=0.96\textwidth]{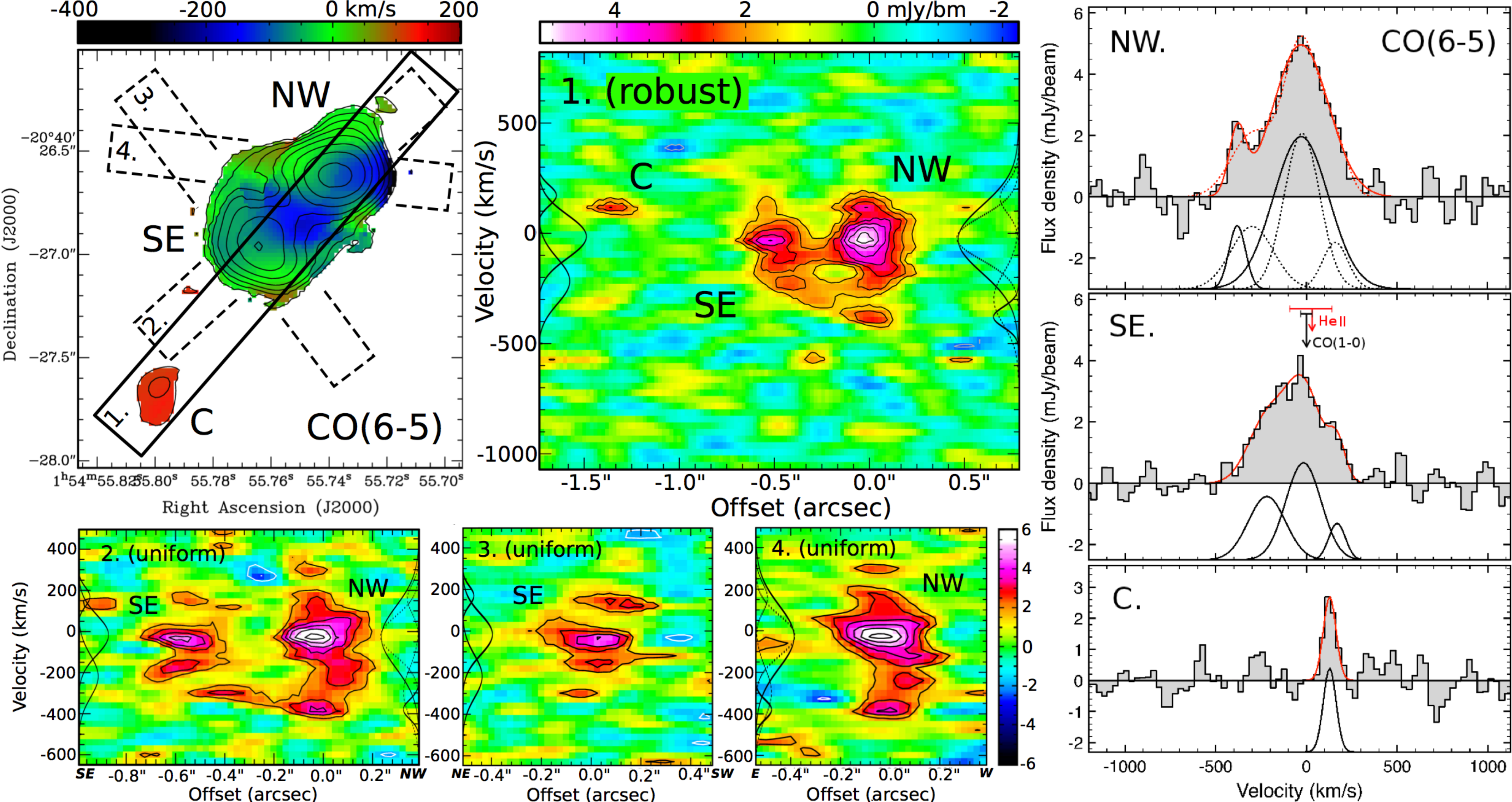}
   \caption{CO(6-5) kinematics. All velocities are with respect to $z$\,=\,1.9212 (EM11). {\sl Top left:} 1$^{\rm st}$ moment velocity map of the CO(6-5) emission, with overlaid contours of the CO(6-5) total intensity at the same levels as those shown in Fig.\,\ref{fig:morphology}. Also indicated are four pseudo-slits along which we extracted the position-velocity plots shown in the middle and bottom (solid for the robust +0.5 weighted data and dashed for the uniform weighted data). The width of the pseudo-slits indicates the spatial resolution. {\sl Top middle:} Position-velocity plot of the robust +0.5 weighted CO(6-5) data set, which highlights the bridge-like structures between NW and SE at v\,$<$\,$-$250\,\kms\ and v\,$\sim$0 \kms\ (PA\,=\,-41$^{\circ}$; direction SE to NW). Contour levels: -4, -3, 3, 4, 5, 6, 7, 8, 9\,$\sigma$, with $\sigma$\,=\,0.54 mJy\,beam$^{-1}$. For guidance, along the y-axis we plot the velocity coverage of the Gaussian model components shown in the spectra on the right. {\sl Bottom:} Position-velocity plots derived from the uniform weighted CO(6-5) data set, which best shows the kinematic details (beamsize 0.27$^{\prime\prime}$\,$\times$\,0.23$^{\prime\prime}$ at PA$_{\rm beam}$\,=\,57$^{\circ}$). The PV-plots are taken along the dashed axes indicated in the top-left plot and all offsets are shown from (north/south)-east to (south/north)-west. Contour levels: -3, -2, 2, 3, 4, 5, 6, 7\,$\sigma$, with $\sigma$\,=\,0.88 mJy\,beam$^{-1}$. {\sl Right:} CO(6-5) spectra of the three components (NW, SE and C). The redshifts $z_{\rm CO(1-0)}$ (EM11) and $z_{\rm He II}$ (EM15), and their uncertainty, are indicated with an arrow\,+\,bar. Also shown are the best-fit model (solid red line) and corresponding individual components (solid black Gaussians lines), as well as an alternative model for NW (dotted lines); see text for details.}
              \label{fig:kinematics}
    \end{figure*}

Figures \ref{fig:morphology} and \ref{fig:kinematics} show our ALMA results. We detect three CO(6-5) components, labeled NW (north-west), SE (south-east), and C (companion). NW and SE are also detected in the 1.2\,mm dust continuum at 245\,GHz. We here discuss the morphology, kinematics, and excitation properties of the components. Values are given in Table\,\ref{tab:results}.

\subsection{Morphology}
\label{sec:morphology}

Figure \ref{fig:morphology} shows that the NW and SE components have a projected separation of $\sim$4 kpc. Their combined {1.2\,mm} flux density ($S_{\rm 1.2\,mm}$ in Table\,\ref{tab:results}) is consistent with the level expected from star formation (SFR\,$\sim$\,3000 M$_{\odot}$\,yr$^{-1}$), with negligible AGN contribution, as modeled by \citet{dro14}. NW and SE are also detected in CO(6-5) emission of cold molecular gas. Component C is found $\sim$10\,kpc in projection towards the south-east and is detected in CO(6-5) but not the dust. Components NW, SE and C all have a counterpart in {\it HST}/NICMOS imaging \citep{pen01}, with NW and SE also detected with {\it HST}/WFPC2 in the UV-restframe (Fig.\,\ref{fig:morphology}). In EM15 we already discussed that both the NICMOS and WFPC2 imaging reveal this double central component.

We argue that NW and SE trace the central kpc-scale region of two galaxies, based on {\sl (1).} their association with individual NICMOS components (which suggests that they contain the bulk of the $\sim$10$^{11.76}$ M$_{\odot}$ stellar mass that was found in the system by \citealt{bre10}); {\sl (2).} the large molecular gas content of both components. Moreover, we will see in Sect.\,\ref{sec:kinematics} that the CO(6-5) kinematics appear inconsistent with the two components being part of a single rotating structure, but that instead they may have their own internal rotation. We also argue that the two galaxies are in the process of merging. This is based on {\sl (1).} their small separation ($\sim$4 kpc), {\sl (2).} the presence of tidal debris on larger scales (Fig.\,\ref{fig:morphology}, see also discussion in EM15); {\sl (3).} the overall high level of star formation in the system (SFR\,$\sim$\,3000 M$_{\odot}$\,yr$^{-1}$). In the remainder of the paper, we therefore refer to the NW and SE components as `nuclei'. C appears to be a small companion galaxy in the NICMOS image, found along a tidal tail. 

An 8.2\,GHz radio continuum image from \citet{pen00radio} reveals two components that hint to a double-lobed, subgalactic-sized radio source associated with NW (Fig.\,\ref{fig:morphology}). This suggests that NW hosts the radio-loud AGN. High resolution radio continuum observations are in progress to verify this.

\subsection{Kinematics}
\label{sec:kinematics}

Figure \ref{fig:kinematics} (left) shows a 1st-moment velocity map of the CO(6-5) emission. Overplotted are the positions of four pseudo-slits along which we extracted the position-velocity (PV) plots shown in Fig.\,\ref{fig:kinematics} (middle + bottom). The main pseudo-slit ($\#$1) shows the robust +0.5 weighted data, which provides the best compromise between resolution and noise level \citep{bri95}. It covers all three CO components (NW, SW and C). The other pseudo-slits show the higher spatial resolution of the noisier uniform weighted data, with slit $\#$2 taken along the axis connecting the NW and SE nucleus, slit $\#$3 along the direction where the total intensity CO emission in SE shows a marginal extension and slit $\#$4 along the axis connecting the NW peak emission with the highest velocity CO feature. Figure \ref{fig:kinematics} (right) shows the CO(6-5) spectra of the three CO components (NW, SE and C). The CO(6-5) emission-line profiles of NW and SE are asymmetric with respect to $z$ derived from CO(1-0) and He\,{\sc II} (EM11, EM15). To better quantify the features seen in the PV plots and emission-line spectra, Fig.\,\ref{fig:kinematics} (right) also shows Gaussian model fits that approximate the complex gas kinematics seen in the PV plots. The best-fit models are derived when using two Gaussian components for NW and three for SE. Based on the CO(6-5) kinematics we determine the following:

\vspace{0.5mm}

\hangindent=0.55cm \hspace{-0.5cm}{\sl 1).} {\sl The CO(6-5) emission peaks at the location and redshift of both NW and SE.} The main Gaussian component in the spectra of NW and SE has a well-defined central velocity of $-$30\,$\pm$\,10 \kms\ for NW and $-$15$\pm$30 \kms\ for SE, which is consistent with $z_{\rm CO(1-0)}$\,=\,1.9212\,$\pm$\,0.0002 (EM11) and $z_{\rm He\,II}$\,=\,1.9214\,$\pm$\,0.0007 (EM15). Our spectral modeling suggests that this main component may have a larger FWHM in NW (360\,$\pm$\,20 \kms) compared to SW (230\,$\pm$\,20 \kms). However, we note that our 2-component fit in NW is not unambiguous, given the limited signal-to-noise and the simplicity of using Gaussian functions to describe the complex gas kinematics. An alternative 3-component model in NW, in which the main Gaussian is similar to that of SE, only leaves residuals that are consistent with the noise, and is thus also acceptable (dotted lines in Fig.\,\ref{fig:kinematics} - right).

\vspace{0.5mm}

\hangindent=0.55cm \hspace{-0.5cm}{\sl 2).} {\sl Most of the CO(6-5) (including gas with velocities up to about $\pm$250 \kms) appears to be distinctively associated with either NW or SE.} The global kinematics of this gas could be consistent with rotation of two individual discs, one for each nucleus. The resolution of our CO(6-5) data is insufficient to securely determine the rotation axis and detailed properties of these putative discs, but their diameters would be $\la$3 kpc, which is smaller than the 4\,kpc separation between NW and SE. The uniform weighted data presented in Fig.\,\ref{fig:kinematics} shows tentative indications that the kinematics of this gas (with |v|\,$\la$\,250\,\kms) could be more complex than simple rotation, but additional observations are needed to further investigate this.



 
\vspace{0.5mm}

\hangindent=0.55cm \hspace{-0.5cm}{\sl 3).} {\sl The highest velocity gas (v\,$\la$\,$-$250\,\kms) appears to form a bridge-like structure between NW and SE.} This emission is the main reason for the asymmetry in the CO(6-5) profiles with respect to $z_{\rm CO(1-0)}$\,=\,1.9212. We therefore argue that this gas feature is not associated with regular rotation. Figure \ref{fig:kinematics} (top-middle) shows that in the robust weighted data there is another faint, bridge-like structure between NW and SE at v$\sim$0\,\kms.

\vspace{0.5mm}

\hangindent=0.53cm \hspace{-0.5cm}{\sl 4).} Component C is redshifted with respect to $z_{\rm CO(1-0)}$. A Gaussian fit yields v=130$\pm$10\,\kms\ and FWHM=80$\pm$20\,\kms.



\subsection{Excitation}
\label{sec:comparison}

The NW nucleus has half the $S_{\rm 1.2\,mm}$ of the SE nucleus, but $\sim$40$\%$ higher $I_{\rm CO(6-5)}$ (Table\,\ref{tab:results}). Unless the dust-to-gas ratio is significantly different for NW and SE, this could indicate that the molecular gas in NW is more highly excited than that in SE. If NW hosts the radio-loud AGN (as argued in Sect.\,\ref{sec:morphology}), this could happen through X-ray radiation from the AGN or shock-excitation by the propagating radio jets \citep[e.g.,][]{ivi12}.

\begin{table}
\caption{Measured and derived physical properties}          
\label{tab:results}      
\centering          
\begin{tabular}{l|c|c|c}     
Region & NW & SE & C. \\
\hline                    
\hline   
R.A.\hspace{3.7mm} (1h\,54m) & 55.738s & 55.763s & 55.799s \\
dec \hspace{4.6mm} (-20$^{\circ}$\,40$^{\prime}$) & 26.61$^{\prime\prime}$ & 26.96$^{\prime\prime}$ & 27.67$^{\prime\prime}$ \\
$S_{\rm 1.2\,mm}$ \hspace{0.0mm} (mJy) & 0.8\,$\pm$\,0.2 & 1.5\,$\pm$\,0.2 & $<$0.3 \\ 
$I_{\rm CO(6-5)}$ (Jy/bm$\cdot$km/s) \hspace{-2mm} & 2.0\,$\pm$\,0.2 & 1.4\,$\pm$\,0.2 & 0.20$\pm$0.05 \\ 
M$_{\rm dust}$ \hspace{2.3mm} (10$^{9}$\,M$_{\odot}$) & 0.5$\pm$0.1 & 0.9$\pm$0.1 & $<$0.24\\     
M$_{\rm H_2}$ \hspace{3.5mm} (10$^{9}$\,M$_{\odot}$) Main$^{\dagger}$ & 14$-$21 & 9\,$\pm$\,2  & 2.2\,$\pm$\,0.5 \\
 \hspace{22.7mm} Wings$^{\ddagger}$ & 1.0$-$7.0 & 2.7$-$5.6 & - \\
\end{tabular}
\flushleft \footnotesize{Errors reflect a conservative 10$\%$ uncertainty in flux calibration plus variation between methods to derive $S_{\rm 1.2\,mm}$ and $I_{\rm CO(6-5)}$ (peak intensity, or integrated intensity from the total intensity map or the line profile).}\\
\footnotesize{$^{\dagger}$ Based on r$_{\rm 6-5}$ = $I_{\rm CO(6-5)}$/$I_{\rm CO(1-0)}$ = 13, $X_{\rm CO} = 0.8$ M$_{\odot}$ (K \kms pc$^{2}$)$^{-1}$.}\\
\footnotesize{$^{\ddagger}$ Based on 17\,$\le$\,r$_{\rm 6-5}$\,$\le$\,36 and the two difference models for NW (see text for details); $X_{\rm CO} = 0.8$ M$_{\odot}$ (K \kms pc$^{2}$)$^{-1}$.}
\end{table}

\begin{figure}
   \centering
   \includegraphics[width=0.44\textwidth]{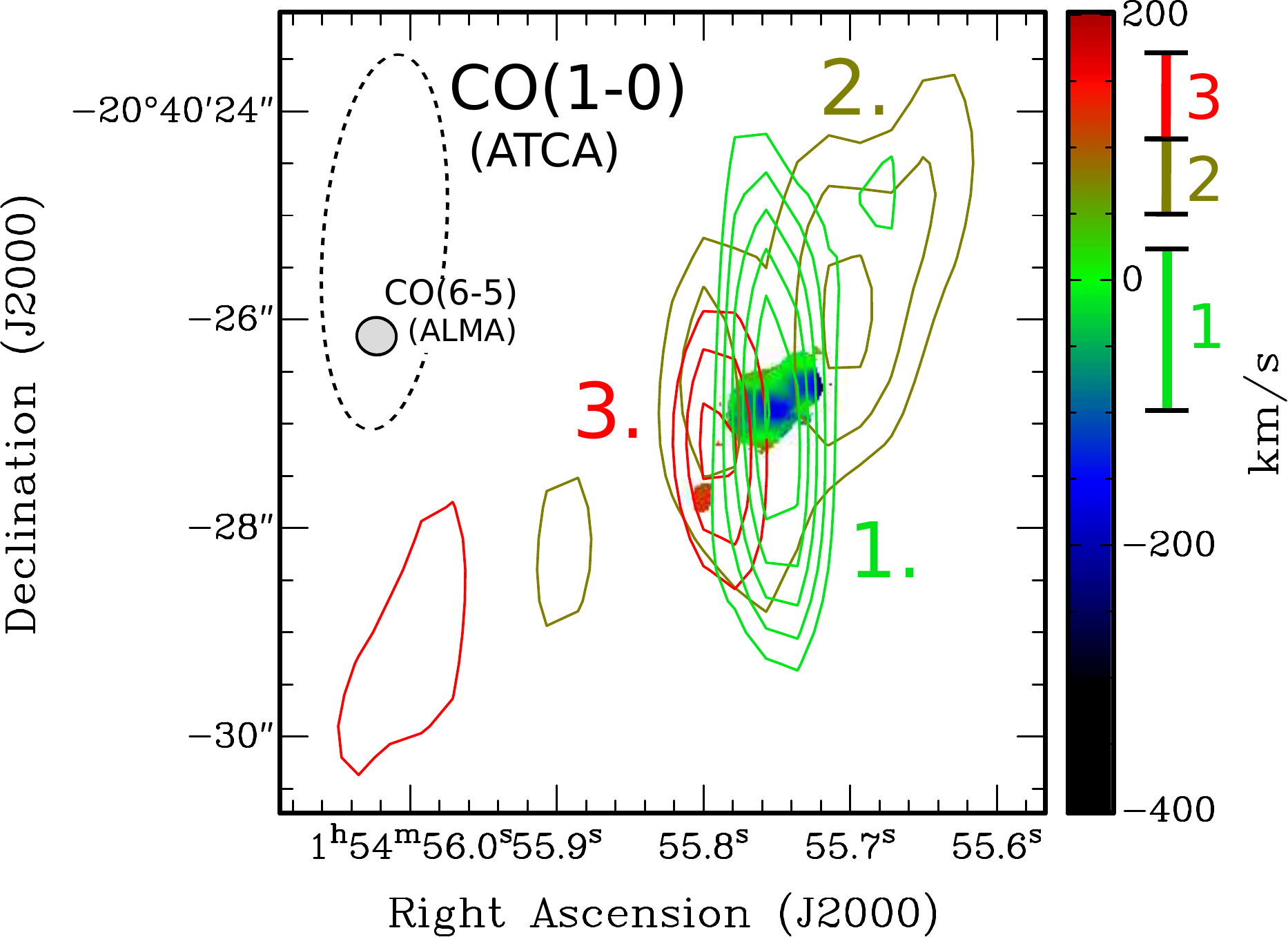}
   \caption{Velocity map of CO(6-5) from Fig.\,\ref{fig:morphology} overlaid onto CO(1-0) contours from ATCA, derived by integrating the CO(1-0) across the velocity ranges -100\,$-$\,+20 \kms\ (light-green $\#$1), +50\,$-$\,+110 \kms\ (dark-green $\#$2) and +110\,$-$\,+170 \kms\ (red $\#$3) (EM15). Levels: 2.8, 3.5, 4.2, 4.9, 5.6, 6.3 $\times$ $\sigma$ ($\sigma$=12 mJy\,beam$^{-1}$\,$\times$\,\kms\ for component $\#$1 and $\sigma$=7.3 mJy\,beam$^{-1}$\,$\times$\,\kms\ for components $\#$2\, and $\#$3). The color-codings of the CO(1-0) contours and CO(6-5) velocity map are closely matched.}
              \label{fig:ATCA}
    \end{figure}

To further investigate the physical conditions and excitation of the gas, Fig.\,\ref{fig:ATCA} and \ref{fig:spectra} compare our ALMA data with lower resolution CO(1-0) data from EM15. Figure \ref{fig:ATCA} shows that the overall kinematics of CO(6-5) and CO(1-0) are in good agreement, but that the CO(1-0) appears to be more widespread than the CO(6-5). A prominent feature seen only in CO(1-0) stretches $\sim$20 kpc north-west of the double nucleus ($\#$2 in Fig.\,\ref{fig:ATCA}). EM15 discuss that this feature is aligned with the radio axis as a possible sign of jet-triggered feedback on tens of kpc scales. In addition, a `red' CO(1-0) component ($\#$3) occurs in between the nucleus and component C, although due to the poor north-south resolution of the ATCA data (see EM15) the offset between this CO(1-0) feature and component C needs to be confirmed. These results suggest that the CO(6-5) traces dense gas in the starforming/AGN regions, while part of the CO(1-0) comes from widespread and likely less dense cold molecular gas (as discussed in EM15).\footnote{When we taper our ALMA data to the same uv-sampling/resolution as the ATCA data, the lower sensitivity only allows us to discard a counterpart to the widespread CO(1-0) in region 2 at the $\sim$2$\sigma$ level. Deeper short-baseline ALMA data are thus needed to draw firm conclusions.}

Figure \ref{fig:spectra} compares the CO(6-5) and CO(1-0) spectra taken against the double (NW+SE) nucleus. The CO(6-5) spectrum was extracted from data smoothed to the same spatial resolution as the CO(1-0) data. From the peak of both spectra, the bulk molecular gas reservoir has an average intensity ratio r$_{\rm 6-5}$\,=\,$I_{\rm CO(6-5)}$/$I_{\rm CO(1-0)}$\,$\sim$\,13, which is consistent with values in the transition region between SMGs and QSOs \citep{wei07}. Higher resolution CO(1-0) data are needed to investigate if r$_{\rm 6-5}$ varies between NW and SE. Interestingly, the blue wing of the CO(6-5) profile in NW and SE has no counterpart in CO(1-0). If we subtract the model of the scaled CO(1-0) spectrum from the CO(6-5) spectrum of the central region, the CO(6-5) residuals reveal this `high-excitation' gas (Fig.\,\ref{fig:spectra}). The 2$\sigma$ CO(1-0) upper limit of this feature in the blue wing is $I_{\rm CO(1-0)}$\,$\le$\,0.065\,mJy\,bm$^{-1}$\,$\times$\,km\,s$^{-1}$ \citep[following][]{emo14}. This constrains the intensity ratio to 17\,$\le$\,r$_{\rm 6-5/1-0}{\rm (blue)}$\,$\le$\,36 (the latter reflects thermalized gas). A fainter high-excitation feature is seen in the residuals of the red wing in Fig.\,\ref{fig:spectra}. These high-excitation features are also identified as the blue and redshifted wings in the spectra of Fig.\,\ref{fig:kinematics}. {\sl This thus shows that the highest velocity gas is also the most highly excited.} Apart from this high-excitation emission, both the CO(1-0) and CO(6-5) profile in Fig.\,\ref{fig:spectra} contain additional excess emission on the redshifted side of the profile, but EM15 show that this is most likely CO emission that leaks in from the molecular gas in or near component C, which is less than one ATCA beam away from the center. 

Figure \ref{fig:spectra} also shows the CO(6-5) total intensity maps across the velocities where both the high-excitation gas is found and the blue/red wing in Fig.\,\ref{fig:kinematics} dominate. The blue and red component are seen opposite of the CO(6-5) peak flux, with a separation between them of 0.17\,$\pm$\,0.05$^{\prime\prime}$ (1.4\,$\pm$\,0.4\,kpc) for NW and 0.20\,$\pm$\,0.06$^{\prime\prime}$ (1.7\,$\pm$\,0.5\,kpc) for SE. This spatial asymmetry confirms that the excess of CO(6-5) in the wings is not because CO(6-5) has a higher velocity dispersion than CO(1-0), but because there is highly excited gas with higher bulk velocities than the gas in the main CO reservoir. 

\section{Discussion}
\label{sec:discussion}

\subsection{Precoalescent merger}

Our ALMA data show that the high-redshift Dragonfly Galaxy is a precoalescent galaxy merger. This merger involves at least two gas- and dust-rich galaxies, and likely a third, smaller gas-rich companion. This is consistent with the claim by \citet{ivi12} that galaxy mergers are ubiquitous among starbursting HzRGs, but the detailed analysis from the combined ALMA and {\it HST} data allowed us to confirm the merger origin, and study the role of the cold gas, with much more confidence than heretofore possible in these systems. While in EM15 (their Sect.\,4.2) we already discussed that the hyper-luminous starburst and powerful radio source were most likely triggered by a gas-rich galaxy merger, our current ALMA results indicate that the triggering of this activity has happened before the parent nuclei fully merged. This is consistent with studies at both low- and high-$z$, which show that the triggering of activity can occur quasi-simultaneously with the merger \citep[e.g.,][]{tad11,car13z4p7}. 

\subsection{Gas and dust masses}

The CO(6-5) emission traces molecular gas in the double nucleus and companion galaxy. From the CO(1-0) emission in the central region, EM15 estimate that the molecular gas mass associated with the double nucleus is M$_{\rm H_2}$ = 2.2\,$\pm$\,0.2 $\times$10$^{10}$ M$_{\odot}$. If we assume the same r$_{\rm 6-5}$ for all three CO(6-5) emitters (Sect.\,\ref{sec:comparison}), then we derive molecular gas masses for NW, SE and C as given in Table \ref{tab:results}. These estimates are based on a conversion factor $X_{\rm CO}$ = M$_{\rm H_2}$/$L'_{\rm CO}$ = 0.8 M$_{\odot}$ (K \kms\ pc$^{2}$)$^{-1}$ found for ultra-luminous IR galaxies \citep{dow98}. Dust masses can be derived following \citet[][]{gil14}, assuming an optically thin case, $\kappa_{\nu} = 4.0(\nu_{\rm rest}/1.2\,{\rm THz})^{-2.0}$ cm$^{2}$\,g$^{-1}$ and $T_{\rm dust} = 45$\,K \citep[][]{ivi12}. M$_{\rm H_2}$/M$_{\rm dust}$, under our assumed r$_{\rm 6-5} \sim 13$, varies by roughly a factor 2 between NW, SE, and C (from Table\,\ref{tab:results}). This implies that the dust-to-gas ratio and/or the gas excitation is substantially different from region to region.

\subsection{High-velocity, high-excitation component: interacting discs or outflows?}

\begin{figure}
   \centering
   \includegraphics[width=0.43\textwidth]{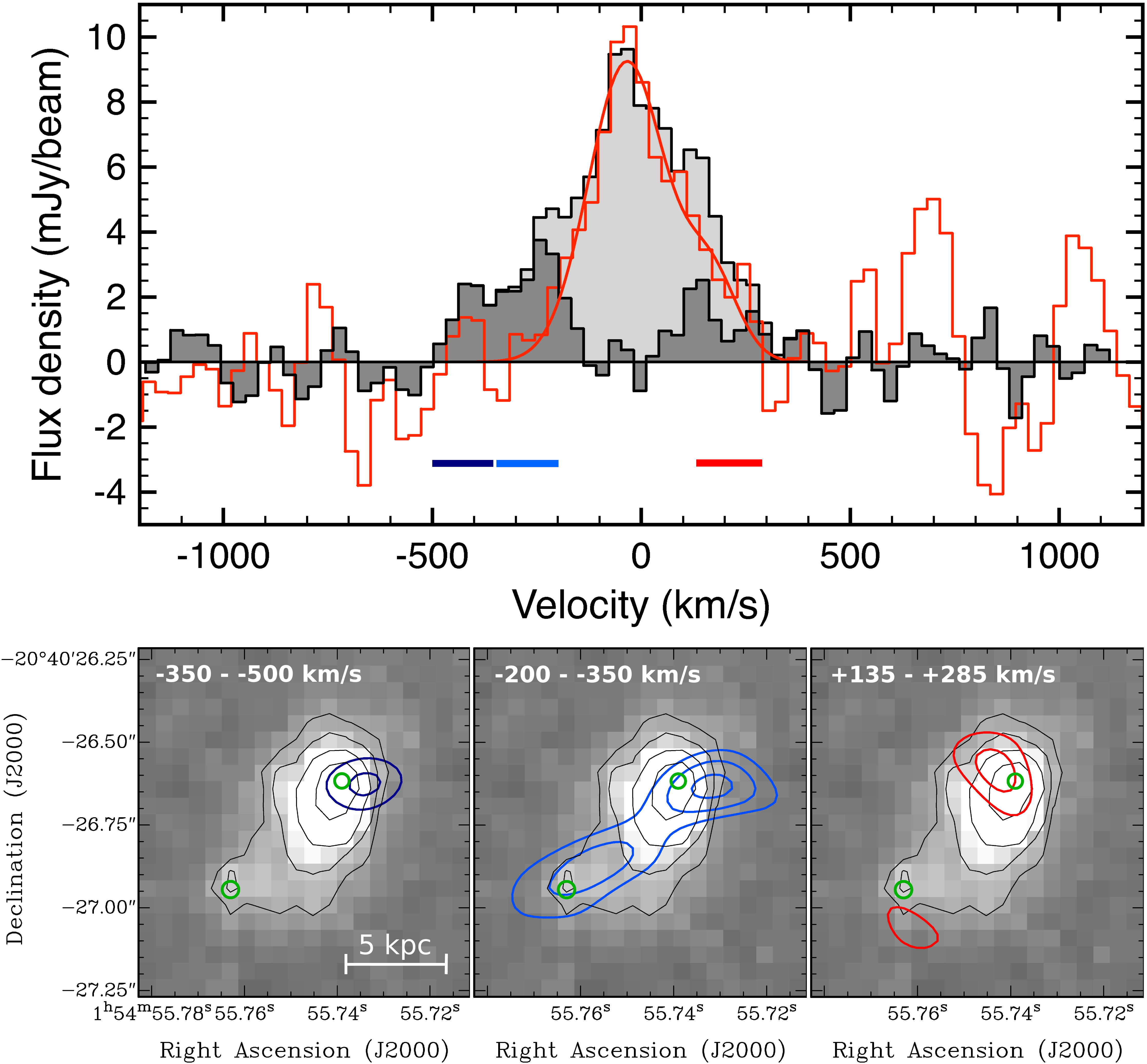}
   \caption{{\sl Top:} Combined CO(6-5) spectrum of NW\,+\,SE, extracted from data smoothed to the resolution of the CO(1-0) data. The red spectrum is the CO(1-0) emission from the same region, scaled up by a factor of 13. The red line is the model fit to the CO(1-0) spectrum (see EM15). The dark-grey area reveals the `high-excitation' residuals when subtracting the CO(1-0) model from the CO(6-5) profile. {\sl Bottom:} Total intensity plots of the velocity ranges where the blue and red wing dominate the CO(6-5) flux, as indicated by the dark-blue, blue and red bars. Contour levels: 2.8, 3.8, 4.8 $\sigma$, with $\sigma = 0.064$ Jy\,beam$^{-1}$\,$\times$\,\kms. The grey-scale image and thin black/white contours show the {\it HST}/WFPC2 image from EM15. The green circles show the location of the CO(6-5) peak emission from the total intensity image of Fig.\,\ref{fig:morphology}, derived by fitting a point-source model to the NW and SE component in the image plane (the size of the circles represents the positional uncertainty of this model-fit).}
              \label{fig:spectra}
    \end{figure}

We also find CO(6-5) emission from molecular gas that has both higher velocities and higher excitation than the bulk of the molecular gas around the systemic velocity. The total mass of this gas is on the order of M$_{\rm H_2}$\,$\sim$\,10$^{10}$ M$_{\odot}$.\footnote{Based on the CO(6-5) signal in the emission-line wings of Fig.\,\ref{fig:kinematics}; see values in Table\,\ref{tab:results}.} Figure \ref{fig:spectra} shows that this high-excitation CO gas is found offset from both the NW and SE nucleus by roughly 1 kpc. A possible scenario is that the bulk of this gas is part of two rotating discs with $\Delta$v\,$\approx$\,$\pm$250\,\kms and R\,$\sim$1\,kpc. The dynamical mass enclosed by each of these two discs would be M$_{\rm dyn}$ $\sim$ 1.5 sin$^{-2}$({\sl i}) $\times$ 10$^{10}$ M$_{\odot}$. When compared to the total stellar mass of the system (M$_{\rm stellar}$ $\sim$ 5.8 $\times$ 10$^{11}$ M$_{\odot}$; \citealt{bre10}), this suggests that either the inclination of at least one of the discs is low (${\sl i} \sim 10 - 20$ degrees), or that we did not trace the full extent of the rotation with the limited spatial resolution and signal-to-noise at which we sample the discs. In this scenario, the highest velocity gas in the CO(6-5) feature between SE and NW ($-$250\,$\la$v\,$\la$$-$450\,\kms) is likely tidal debris from an ongoing interaction between these discs. This tail-like feature has $I_{\rm CO(6-5)} \sim 0.54$ Jy\,beam\,$\times$\,\kms, which corresponds to a molecular gas mass of M$_{\rm H2} \sim 2 - 5 \times 10^{9}$ M$_{\odot}$ (assuming 17\,$\le$\,r$_{\rm 6-5}$\,$\le$\,36 and $X_{\rm CO}$ = 0.8 M$_{\odot}$ (K \kms\ pc$^{2}$)$^{-1}$). The scenario of two interacting discs would be in agreement with numerical simulations of galaxy collisions at $z$\,$\sim$\,2, which show that mergers involving cold and clumpy gas discs have chaotic velocity fields with turbulent gas kinematics while reaching star formation rates of a few 1000\,M$_{\odot}$\,yr$^{-1}$ \citep[][see also, e.g., \citealt{dim07}]{bou11}. One puzzling fact with this scenario is that the gas in the outer parts of the discs has a higher excitation than the gas in the center, despite the likely presence of a central starburst and, for at least one of the nuclei, also an AGN.

\vspace{0mm}

Alternatively, all of the high-excitation, high-velocity gas in the wings of the CO(6-5) profile could be molecular material that is driven out of the two central nuclei, either through gravitational forces or through outflows induced by the starburst or AGN. Bi-conical molecular outflows have been seen in nearby active galaxies \citep[e.g.,][]{ala11,emo14_NGC3256,sak14,tad14}, while massive outflows of ionized gas have been found in HzRGs \citep{nes08}. These ionized outflows show velocities that are at most a factor 1.5\,$-$\,2 higher than those of CO(6-5), although they are more turbulent (FWHM$\sim$1000 \kms) and extend out to $\sim$10 kpc scales. A similar large-scale outflow of [C\,{\sc II}] gas was recently discovered in the mm-regime in a $z$\,$\sim$\,6 QSO \citep{cic15}. To explore whether the outflow scenario is feasible, we have to assess the energetics involved. The combined bulk and turbulent kinetic energy of the outflowing high-excitation molecular gas would be $E$ = $\frac{1}{2}$M[v/sin({\sl i}$_{\rm outfl}$)]$^{2}$ + $\frac{3}{2}$M($\frac{\rm FWHM}{2.36}$)$^{2}$ $\sim$ 1\,$\times$\,10$^{58}$ erg (assuming {\sl i}$_{\rm outfl}$\,=\,90$^{\circ}$, M$_{\rm H_2}$\,$\sim$\,10$^{10}$ M$_{\odot}$, v\,$\sim$\,250\,\kms\ and FWHM\,$\sim$\,300\,\kms). Following \citet{hec93}, we estimate that the starburst \citep[$L_{\rm 8-1000}\mu$m \,$\sim$\,1.8\,$\times$\,10$^{13}$\,L$_{\odot}$;][]{dro14} releases a kinetic energy of $E_{\rm kin}$\,$\sim$\,1\,$\times$\,10$^{59}$ ergs over the 3 Myr time-scale that the outflow requires to cover 1 kpc. Although these are merely order-of-magnitude calculations, additional energy from supernova and the AGN is released into the system, which indicates that the current level of activity can indeed produce enough energy to drive a putative molecular outflow. Additional observations are required to investigate this scenario.

The question is whether the gas at the highest velocities can potentially escape the merger system? If we consider the lower mass limit of the system by only taking into account the mass of the stars \citep[M$_{*}$\,$\sim$\,5.8$\times$10$^{11}$\,M$_{\odot}$;][]{bre10} and assume that the bulk of this mass is concentrated in the merging nuclei, then the corresponding escape velocity at a distance of 4\,kpc (i.e., the distance between the merging nuclei) is of the order of $\sim$1000 \kms. This means that, at the |v|\,$\la$\,500 \kms\ that we observe, the molecular gas will likely remain gravitationally bound to the system as a potential fuel reservoir for future star formation.

Our ALMA data thus suggest that molecular gas is being redistributed within the merger system, but it remains unclear as to what is the total mass currently involved and what is the main driving mechanism (i.e., tidal forces between two rotating discs or outflows). When only considering the high-velocity tail ($-$250\,$\la$\,v\,$\la$$-$450 \kms), molecular gas is being redistributed at a minimum rate of \.{M}\,=M$_{\rm H2}$\,$\times$\,v$_{\rm ejecta}$\,$\times$\,R$^{-1}$\,\,$\sim$\,1200\,$\pm$\,500 M$_{\odot}$\,yr$^{-1}$ (assuming M$_{\rm H2} \sim 2 - 5 \times 10^{9}$ M$_{\odot}$, v$_{\rm ejecta}$\,$\sim$\,350\,\kms\ and R\,$\sim$\,1\,kpc). In the scenario that all of the high-excitation gas is being ejected, we can place an upper limit on the redistribution rate of \.{M}\,=M$_{\rm H2}$\,$\times$\,v$_{\rm max}$\,$\times$\,R$^{-1}$\,\,$\sim$\,2900\,$\pm$\,1600 M$_{\odot}$\,yr$^{-1}$ (following \citealt{arr14} and assuming a conservative v$_{\rm max}$\,=\,$|\Delta {\rm v}|$+FWHM/2\,$\sim$\,350\,\kms, R$_{\rm outfl}$\,$\sim$\,1 kpc and M$_{\rm H2}$ as per Table\,\ref{tab:results}). This is similar to the total SFR\,$\sim$\,3000\,$\pm$\,800 M$_{\odot}$\,yr$^{-1}$ in the Dragonfly Galaxy \citep{dro14}.\footnote{The error is based on the uncertainty in $L_{\rm IR}$ \citep{dro14} combined with variations in the constants used in the Kennicutt law \citep{ken98}, as discussed by \citep{cal12}, depending on IMF and specific star formation time.} Our results thus suggest that a significant fraction of the fuel for the ongoing star formation is rapidly displaced, and possibly even transferred from one nucleus to the other. In this respect, the time-scale for the gas to cover the 4\,kpc distance between the nuclei at v\,$\sim$350\,\kms\ is roughly 11 Myr (not taking into account possible geometric effects). This is on the order of the $\sim$7 Myr minimum gas depletion time due to star formation in the central region (EM15). This suggests that in a major merger like the Dragonfly Galaxy, gas can be redistributed on roughly the same time-scale as that it is being depleted by star formation, which may result in bursty star-formation episodes.

\section{Conclusions}
\label{sec:conclusions}

Concluding, our ALMA results show that the Dragonfly Galaxy is undergoing a short but crucial epoch of peak activity, triggered by a precoalescent gas-rich merger. During this epoch, the molecular gas is rapidly consumed by vigorous star formation. At the same time, a significant fraction of the gas is displaced and apparently excited as a result of tidal forces or possible outflows. The rate at which the gas is redistributed may approach the rate at which it is depleted by star formation. This suggests that, besides star formation efficiency, dynamical effects can also be important in the early evolution of massive galaxies. ALMA in its full potential will be able to further investigate this. The merger processes in the Dragonfly Galaxy are likely scaled-up versions of those observed in low-$z$ powerful radio galaxies, and will help us understand the early stages in the evolution of high-$z$ radio galaxies into current-day massive ellipticals.

\begin{acknowledgements}
We thank the ALMA staff for their dedicated work. BE is grateful that the research leading to these results has been funded by the European Union 7$^{th}$ Framework Programme (FP7-PEOPLE-2013-IEF) under grant 624351. MVM's work was funded with support from the Spanish Ministerio de Econom\'{i}a y Competitividad through grant AYA 2012-32295. NS is recipient of an ARC Future Fellowship. We used the following ALMA data: 2013.1.00521.S. ALMA is a partnership of ESO (representing its member states), NSF (USA) and NINS (Japan), together with NRC (Canada) and NSC and ASIAA (Taiwan), in cooperation with the Republic of Chile. The Joint ALMA Observatory is operated by ESO, AUI/NRAO and NAOJ.
\end{acknowledgements}



\end{document}